\documentclass[12pt]{article}
\usepackage{amsmath,amssymb,amsthm,bbm,natbib,color,hyperref,caption,graphicx,subcaption,afterpage}
\usepackage[margin=1in]{geometry}
\usepackage[hang,flushmargin]{footmisc}
\usepackage[table,xcdraw]{xcolor}
\usepackage{tikz}
\usepackage{tkz-euclide}

\usepackage{algcompatible} 
\usepackage[]{algorithm} 

\newtheorem{theorem}{Theorem}
\newtheorem{proposition}{Proposition}

\newcommand{\R}{\mathbb{R}}

\newcommand{\EE}[1]{\mathbb{E}\left[{#1}\right]}

\newcommand{\Pp}[2]{\mathbb{P}_{{#1}}\left\{{#2}\right\}}

\newcommand{\Ppst}[3]{\mathbb{P}_{{#1}}\left\{{#2}\  \middle| \ {#3}\right\}}

\newcommand{\iidsim}{\stackrel{\textnormal{iid}}{\sim}}
\newcommand{\indsim}{\stackrel{\perp\!\!\!\!\perp}{\sim}}

\graphicspath{{./figs/}}

\title{Selective inference for clustering \\ with unknown variance}
\author{Young-Joo Yun\thanks{Department of Statistics, University of Wisconsin---Madison} \and
 Rina Foygel Barber\thanks{Department of Statistics, University of Chicago}}

\begin{document}
\maketitle

\begin{abstract}
 In many modern statistical problems, the limited available data must be used both to 
develop the hypotheses to test, and to test these hypotheses---that is, both for exploratory and confirmatory data analysis.
Reusing the same dataset for both exploration and testing can lead to massive selection bias, leading
to many false discoveries. Selective inference is a framework that allows for performing valid inference
even when the same data is reused for exploration and testing. In this work, we are interested in the problem of selective inference
for data clustering, where a clustering procedure is used to hypothesize a separation of the data points into a collection of 
subgroups, and we then wish to test whether these data-dependent clusters in fact represent meaningful differences within the data.
Recent work by \citet{gao2020selective} provides a framework for doing selective inference for this setting,
 where a hierarchical clustering algorithm is used for producing the cluster assignments,
which was then extended to k-means clustering by \citet{chen2022selective}. Both these
works rely on assuming a known covariance structure for the data, but in practice, the noise level needs to be estimated---and this is 
particularly challenging when the true cluster structure is unknown. In our work, we extend 
this work to the setting of noise with unknown variance, and provide a selective inference method for this more general setting.
Empirical results show that our new method is better able to maintain high power while controlling Type I error
when the true noise level is unknown.
\end{abstract}

\section{Introduction}\label{sec:intro}
Data clustering is a popular method for summarizing trends in unlabeled data,
and is a powerful tool for gaining understanding and interpretability in large datasets. However, it is well known that clustering can easily lead to false discoveries,
in the sense that data from a single source (one true cluster) can be falsely 
separated into multiple clusters. A core challenge in addressing this issue is that of
{\em selective inference}---the problem of performing inference on a hypothesis that was
developed using the same dataset. In-depth motivations for selective inference can be found in  \citet{taylor2015statistical} and \citet{benjamini2020selective}, where the latter illustrates its importance from the perspective of replicability of experimental results. There has been extensive work on selective inference in supervised settings, such as the work by \citet{fithian2014optimal} and  \citet{loftus2015selective} that provide selective inference frameworks for doing valid inference after model selection, but less is known about the unsupervised setting, which is the context for clustering. Some popular clustering methods include hierarchical clustering (\citet{murtagh2012algorithms}), k-means clustering (\citet{steinley2006k}), decision tree clustering (\citet{liu2000clustering}), and spectral clustering (\citet{von2007tutorial}).

Concretely, for the clustering problem,
if the $n$ data points $X_1,\dots,X_n\in\R^q$ are partitioned into
clusters $C_1\cup\dots\cup C_K = [n]:=\{1,\dots,n\}$, how can we determine whether
the ``discovered'' clusters $C_k$ and $C_{k'}$ are genuinely different using
the observed data (the $X_i$'s for $i\in C_k$, and for $i\in C_{k'}$), when these
same data values were used to choose the clusters themselves? 

To address this problem, \citet{zhang2019valid} propose a method based on data splitting, where the hyperplane separating two clusters is estimated using a portion of the data, and the fitted hyperplane, instead of the clustering algorithm, is used on the rest of the data for generating cluster assignments. They then condition on the selection event---the event where the hyperplane is chosen---to account for the data dependency in the hypothesis test. \citet{gao2020selective} provide an alternative solution to this problem, where they account for the clustering event by directly conditioning on it, specifically for the hierarchical clustering algorithm, and a recent work of \citet{chen2022selective} extends their work to the k-means clustering algorithm. Relatedly, \citet{hivert2022post} propose a set of valid inference procedures for three different null hypotheses testing whether two clusters estimated by a clustering algorithm are truly different, where one of the proposed procedures is an extension of the work of \citet{gao2020selective} to testing whether a single feature plays a role in distinguishing the two clusters. There have also been relevant work on data with structural assumptions. For example, the work of \citet{watanabe2021selective} provides a method for
doing inference on a data matrix represented by a latent block model after choosing the cluster membership of each entry of the data matrix using the same data matrix, and conditions on this selection event to do a valid inference.

The aforementioned work of \citet{gao2020selective} offers an elegant solution to this problem,
providing a framework for performing selective inference to test the null hypothesis that states \begin{equation}\label{eqn:null_original}H_0(C_k,C_{k'}) : \ \frac{1}{|C_k|} \sum_{i\in C_k}\mu_i = \frac{1}{|C_{k'}|} \sum_{i\in C_{k'}}\mu_i \end{equation}
for each pair $k\neq k'$, where $\mu_i = \EE{X_i}$ is the true mean of the $i$-th data
point---in other words, is the mean of cluster $C_k$ equal to the mean of cluster $C_{k'}$? Note that this hypothesis is indeed data-dependent, since the clusters $C_k$ and $C_{k'}$
are chosen based on the observed data, and therefore testing this null must 
account for this data dependence. In  \citet{gao2020selective}'s work, it is assumed that the data
is distributed as 
\begin{equation}\label{eqn:data_gen}
X_i\indsim \mathcal{N}(\mu_i,\sigma^2\mathbf{I}_q)
\end{equation}
for $i\in[n]$, where the means $\mu_i\in\R^q$ are unknown
while the (shared) variance $\sigma^2>0$ is known. In practice, however, $\sigma$ would often need
to be estimated from the data, and this poses a particular challenge in the setting of this clustering problem. Without knowing the true cluster structure of the data (since of course, this is exactly the question we are aiming to test), it is difficult to obtain a reliable estimate of $\sigma$--indeed, we will see shortly that many natural options lead to either substantial power loss or substantial loss of the Type I error control. This motivates the need for the more general model that avoids the need to estimate the true variance. In this work, we propose a method that avoids this obstacle, by allowing for an unknown variance $\sigma^2$ (or more generally, an unknown structured covariance matrix), while guaranteeing Type I error control and maintaining high power. 

The remainder of this paper is organized as follows. 
In Section~\ref{sec:background}, we review the selective inference framework
developed by \citet{gao2020selective} for the setting where $\sigma$ is known and discuss motivations for allowing $\sigma$ to be unknown.
In Section~\ref{sec:method}, we present our new method for performing inference on clustering in the setting of an unknown $\sigma$ (with proofs
deferred to the Appendix).
Empirical results are presented in Section~\ref{sec:empirical} to demonstrate the performance
of the new method and compare against the existing framework. Finally, we 
conclude with a discussion and some open questions in Section~\ref{sec:summary}.

\section{Background: the known variance case}\label{sec:background}

In this section, we will first give a brief overview of the selective inference method developed in  \citet{gao2020selective}'s
work, and discuss some of the challenges posed by unknown variance $\sigma^2$.

\subsection{\citet{gao2020selective}'s Method}\label{sec:background_gao_method}

Consider clusters $C_k,C_{k'}$, which are two disjoint subsets of $[n]$. If 
these clusters were chosen {\em ahead of time}---that is, independently of the data---then it would
be simple to test the null hypothesis $H_0(C_k,C_{k'})$ defined in~\eqref{eqn:null_original}---specifically,
we would naturally use the test statistic
\[ \frac{1}{|C_k|} \sum_{i\in C_k} X_i-\frac{1}{|C_{k'}|} \sum_{i\in C_{k'}} X_i = X^\top v  \textnormal{ where } v:=\frac{\mathbf{1}_{C_k}}{|C_k|} -\frac{\mathbf{1}_{C_{k'}}}{|C_{k'}|}.\] 
Here $X\in\R^{n\times q}$ is the matrix of observed data with $i$-th row $X_i\in\R^q$, and 
where, for a subset $C\subseteq [n]$,  $\mathbf{1}_C\in \R^n$ represents the vector with $i$th entry equal to $1$ for each $i\in C$ and $0$ for $i\not\in C$. This test statistic
follows a mean-zero normal distribution under the null hypothesis $H_0(C_k,C_{k'})$,
and so its norm follows a rescaled $\chi$ distributed under the null,
\[\|X^\top v\|_2\stackrel{H_0(C_k,C_{k'})}{\sim}\sigma\left(\frac{1}{|C_k|} + \frac{1}{|C_{k'}|}\right)^{1/2} \cdot \chi_q.\]

However, since the clusters were chosen in a data-dependent way,
this distribution is not the correct null distribution for $\|X^\top v\|_2$.
To address this, we can rewrite $X$ as
\[X = \mathcal{P}_v X + \mathcal{P}_v^\perp X =\frac{\|X^\top v\|_2}{\|v\|_2} \cdot \frac{vv^\top X}{\|vv^\top X\|_F} +\mathcal{P}_v^\perp X,\]
which decomposes $X$ into components lying in the span of $v$ and its orthogonal complement,
with $\mathcal{P}_v =\frac{vv^\top}{\|v\|^2_2}$ denoting the projection matrix that projects
to the span of $v$, and  $\mathcal{P}_v^\perp = \mathbf{I}_n-\frac{vv^\top}{\|v\|^2_2}$ projecting to its orthogonal complement, and where $\|\cdot\|_2$ denotes the Euclidean norm and $\|\cdot\|_F$ the Frobenius norm.
\citet{gao2020selective}'s insight into handling the data-dependent cluster selection
is to condition on the normalized matrix $ \frac{vv^\top X}{\|vv^\top X\|_F} $ and the orthogonal projection $\mathcal{P}_v^\perp X$,
so that only the test statistic $\|X^\top v\|_2 $ remains unknown, and moreover to condition
on the range of values of $\|X^\top v\|_2 $ that agree with the clustering selection. Specifically,
defining
\begin{equation}\label{eqn:decomp_Gao}x(\phi) = \frac{\phi}{\|v\|_2} \cdot   \frac{vv^\top X}{\|vv^\top X\|_F} +\mathcal{P}_v^\perp X,\end{equation}
let
\[\mathcal{S} = \left\{\phi > 0 : \textnormal{Cluster}(X) = \textnormal{Cluster}(x(\phi))\right\},\]
where $\textnormal{Cluster}(\cdot)$ refers to the outcome of the clustering procedure.
In other words, $\mathcal{S}$ contains all values of $\phi$ for which the same clustering outcome
would have been obtained, if we plug in $\phi$ in place of the observed test statistic value $\|X^\top v\|_2$. Their main result establishes that, even given the data-dependent clustering procedure,
the re-scaled $\chi$ distribution is the correct null distribution once truncated to this set $\mathcal{S}$.
\begin{theorem}[{\citet[Theorem 1]{gao2020selective}}]\label{thm:gao}
Let $X_i\indsim \mathcal{N}(\mu_i,\sigma^2\mathbf{I}_q)$ where $\sigma$ is known, and let $v$ be defined as above.
Then, conditional on $\textnormal{Cluster}(X)$, $ \frac{vv^\top X}{\|vv^\top X\|_F} $, and $\mathcal{P}_v^\perp X$,
under the null hypothesis $H_0(C_k,C_{k'})$ the test statistic $\|X^\top v\|_2$ follows a truncated
rescaled $\chi$ distribution,
$\sigma\left(\frac{1}{|C_k|} + \frac{1}{|C_{k'}|}\right)^{1/2} \cdot \chi_q$ truncated to $\mathcal{S}$.
In particular, the p-value
\[P = 1-F_{\chi_q}\left(\|X^\top v\|_2 ; \sigma\left(\frac{1}{|C_k|} + \frac{1}{|C_{k'}|}\right)^{1/2}, \mathcal{S}\right)\]
is uniformly distributed under $H_0(C_k,C_{k'})$, where $F_{\chi_q}(\cdot;c,\mathcal{S})$ is the
CDF of a $c\cdot\chi_q$ random variable truncated to the set $\mathcal{S}$.
\end{theorem}

\citet{gao2020selective} provide an algorithm for exactly computing the set $\mathcal{S}$ for the hierarchical clustering algorithm with linkages for which the exact computation of this set is tractable, along with an implementation of an importance sampling algorithm for clustering algorithms where this set cannot be efficiently computed.

 It can be seen in Theorem \ref{thm:gao} that \citet{gao2020selective} condition on $\textnormal{Cluster}(X)$, which is the entire outcome of the clustering algorithm 
 and contains information about all $K$ estimated clusters, as opposed to conditioning only on the event $C_k,C_{k'}\in \textnormal{Cluster}(X)$, where $C_k$ and $C_{k'}$ are the two clusters of interest. 
We might ask whether it is necessary to condition on this additional information, or whether this might lead to loss of power.
Indeed, this type of approach is common in the selective inference literature---in order to be able to perform selective inference,
we may need to condition on additional information for statistical reasons (i.e., to avoid nuisance parameters) and/or computational reasons;
see, for instance, \citet[Section 3.2.4]{fithian2015topics}, \citet[Section 5.2]{lee2016exact} for more discussion.

\subsection{Challenges in estimating $\sigma$}\label{sec:background_challenges}
We next discuss motivations for allowing $\sigma$ to be unknown. Continuing the discussion earlier on the difficulty of estimating $\sigma$ from the data, we consider a simple scenario where we are aiming to determine whether data points $X_1,\dots,X_n$
arise from a single cluster or from two clusters. To test this, we would  choose
a data-dependent clustering $[n]=C_1\cup C_2$, and would now need to estimate $\sigma$ in 
order to run  \citet{gao2020selective}'s test.

\begin{figure}[t]
    \centering
    {{\includegraphics[width=15cm]{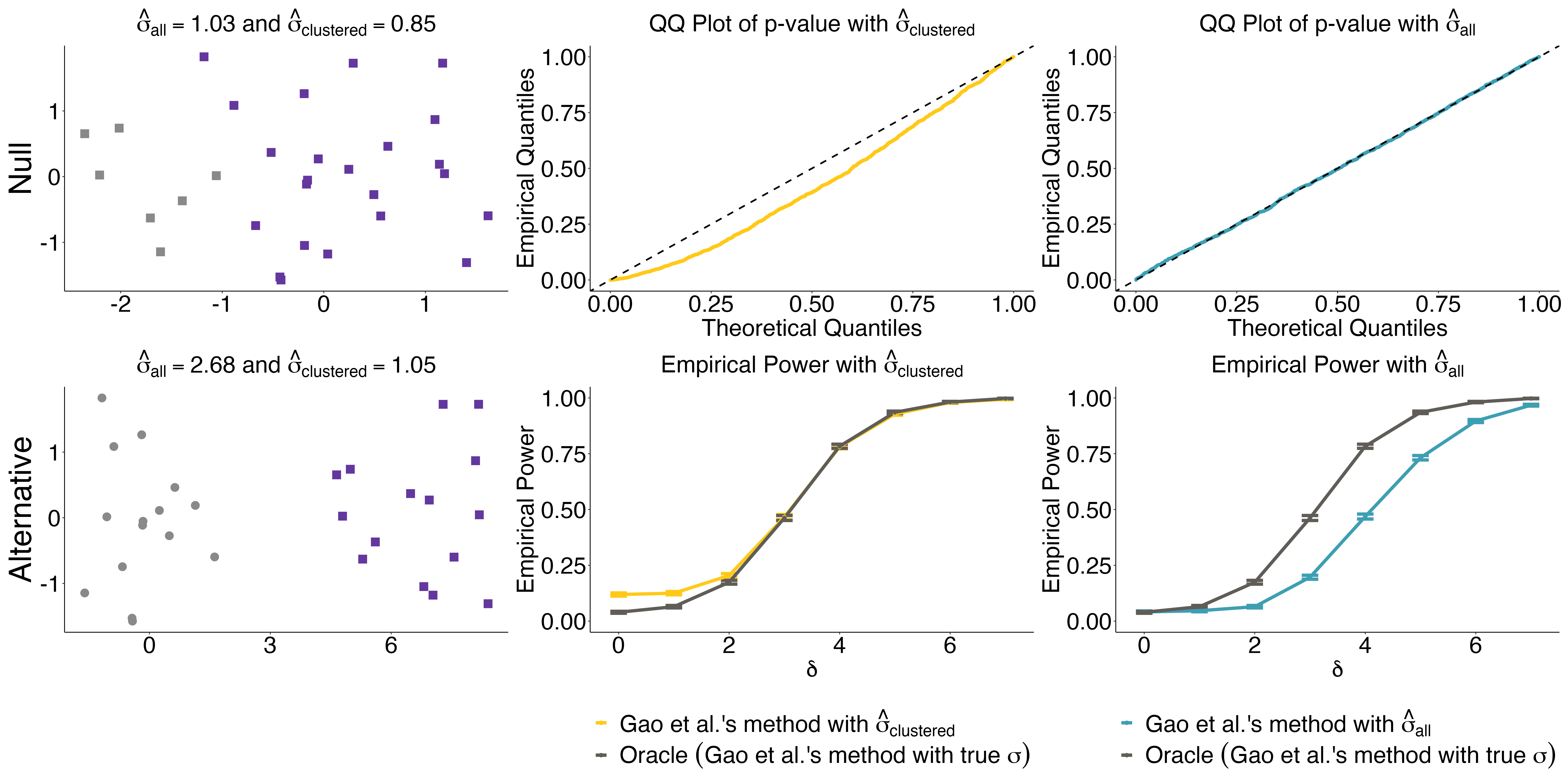}}}
    \caption{The top row shows results under the null, and the bottom row shows results under the alternative.
    In each row, the left plot shows one draw of the data, along with the estimated values
     $\hat\sigma_{\textnormal{all}}$ and $\hat\sigma_{\textnormal{clustered}}$, while the middle and right plots show results for 
      \citet{gao2020selective}'s method applied with $\hat\sigma_{\textnormal{clustered}}$ or with $\hat\sigma_{\textnormal{all}}$, respectively.
      (See Section~\ref{sec:background_challenges} for discussion.)}\label{fig_motivation}
\end{figure}

\begin{itemize}
\item Suppose we estimate the variance by using the within-cluster means, for instance,
\[\hat\sigma_{\textnormal{clustered}}^2 = \frac{\sum_{i\in C_1}\|X_i-\bar{X}_{C_1}\|^2_2 + \sum_{i\in C_2} \|X_i-\bar{X}_{C_2}\|^2_2}{(n-2)q},\]
where $\bar{X}_{C_k}$ is the sample mean in cluster $C_k$. With this choice, we might
substantially underestimate the variance if the true data distribution only has a single cluster. The middle column of Figure \ref{fig_motivation}
demonstrates this problem in practice---we can see that, when the null $H_0(C_1,C_2)$ is true,
the variance may sometimes be vastly underestimated and, as a result,
the empirical distribution of the p-value is far from uniform, which would lead to false positives.
\item Alternatively, we might take a more conservative estimate of variance by treating
the data as a single cluster, e.g.,
\[\hat\sigma_{\textnormal{all}}^2 = \frac{\sum_{i\in [n]} \|X_i-\bar{X}_{[n]}\|^2_2 }{(n-1)q}.\]
Indeed, this is the estimator proposed in \citet[Section S3]{gao2020selective}, and they prove theoretically
that, as this is asymptotically an over-estimate of $\sigma^2$, Type I error control is guaranteed.
However, this choice can lead to a substantially over-conservative test, as demonstrated 
in the right column of Figure~\ref{fig_motivation}---if the true data distribution arises from two clusters, this estimate can massively over-estimate
$\sigma^2$ leading to a large loss of power.\footnote{
In sparse high-dimensional settings, \citet{chen2022selective} propose using an alternative definition based on the median rather than the mean,
$\hat{\sigma}^2_{\textnormal{all--med}}=\textnormal{Median}\{\tilde{X}^2_{ij}:i\in [n], j\in [q]\}/\textnormal{Median}(\chi^2_1)$, where $\tilde{X}_{ij} = X_{ij} - \textnormal{Median}\{X_{i'j}: i'\in[n]\}$.
They observe that this alternative estimator is less conservative if cluster mean differences are sparse.
Since we work in a low-dimensional setting, we do not compare to this alternative.
}
\end{itemize}
See Section~\ref{sec:empirical} for details on these simulations.

Thus, in Figure~\ref{fig_motivation}, we clearly see a tradeoff between Type I error control and power. When using the cluster-wise estimate $\hat\sigma_{\textnormal{clustered}}$, we see that power is high under the alternative, with the empirical power being as high as the case where the true $\sigma$ is used, but Type I error control is lost under the null. On the other hand, when using the estimate $\hat\sigma_{\textnormal{all}}$ that treats the entire dataset as a single cluster, we see that it controls Type I error under the null but incurs a loss in power under the alternative.

To avoid this tradeoff, in this work we propose a selective inference procedure for the clustering
problem that can handle an unknown variance $\sigma^2>0$. This more general model resolves the issue---the p-value distribution is uniform when the data is generated from a single cluster, without sacrificing too much power in the scenario where the data is instead generated from distinct clusters.

\section{Proposed method: the unknown variance case}\label{sec:method}

We now introduce our proposed method for the setting where the variance is unknown. In this new setting, we assume that the data
is distributed as $X_i\indsim \mathcal{N}(\mu_i,\sigma^2\mathbf{I}_q)$, where the means $\mu_i\in\R^q$, as well as the (shared) variance $\sigma^2>0$, are unknown.

Recall the null hypothesis
\[H_0(C_k,C_{k'}) : \ \frac{1}{|C_k|} \sum_{i\in C_k}\mu_i = \frac{1}{|C_{k'}|} \sum_{i\in C_{k'}}\mu_i\]
in \citet{gao2020selective}, and the corresponding test statistic $\|X^\top v\|_2$. Unfortunately, in our new setting where $\sigma$ is unknown, the distribution of this test statistic cannot be computed.
With unknown $\sigma$, a typical approach would be to construct a data dependent estimator of $\sigma$ to result in a test statistic of the form $\|X^\top v\|^2_2 / \textnormal{(estimate of $\sigma^2$)}$, which would  follow an F distribution (after rescaling the numerator appropriately). However, this is impossible to do in our setting under the broad null hypothesis $H_0$.
This is due to the fact that, to estimate $\sigma$, we would need to examine the within-cluster variability, i.e., quantities of the form $\sum_{i\in C_k}\|X_i - \bar{X}_{C_k}\|^2_2$ for estimated clusters $k=1,\dots,K$; however, these quantities follow a \emph{noncentral} $\chi^2$ distribution when each cluster $C_k$ might contain \emph{different} means $\mu_i$, following the distribution $\sigma^{-2}\sum_{i\in C_k}\|X_i - \bar{X}_{C_k}\|^2_2 \sim \chi^2_{q(|C_k|-1)}(\sigma^{-2}\sum_{i\in C_k}\|\mu_i - \bar{\mu}_{C_k}\|^2_2)$.  In other words, any nonzero differences in means within a cluster will lead to nuisance parameters, arising from within-cluster differences in means, that make it impossible to handle unknown variance.

To avoid these nuisance parameters, we restrict to a stronger null hypothesis,
\begin{equation}\label{eqn:null_new}H_0'(C_k,C_{k'}) : \ \mu_i = \mu_{i'} \ \forall \ i,i'\in C_k\cup C_{k'}.\end{equation} 
In other words, $H_0'$ assumes that {\it each data point} in clusters $C_k$ and $C_{k'}$ has the same mean,
while $H_0$ makes the weaker assumption that the {\it sample mean} of data points in  cluster $C_k$ and in cluster $C_{k'}$ have the same mean. We can equivalently rewrite $H_0'(C_k,C_{k'})$ as 
\begin{equation}\label{eqn:null_new_equiv}H_0'(C_k,C_{k'}):\left(\mathbf{I}_n-\frac{ww^\top}{\|w\|^2}-\underset{i\in[n]\backslash (C_k\cup C_{k'})}{\sum}e_ie_i^\top\right)\mu=0\textnormal{ where }w:=\frac{\mathbf{1}_{C_k\cup C_{k'}}}{|C_k\cup C_{k'}|}.\end{equation}

\subsection{Decomposition of $X$}
\begin{figure}[hbtp]
    \centering
    {{\includegraphics[width=15cm]{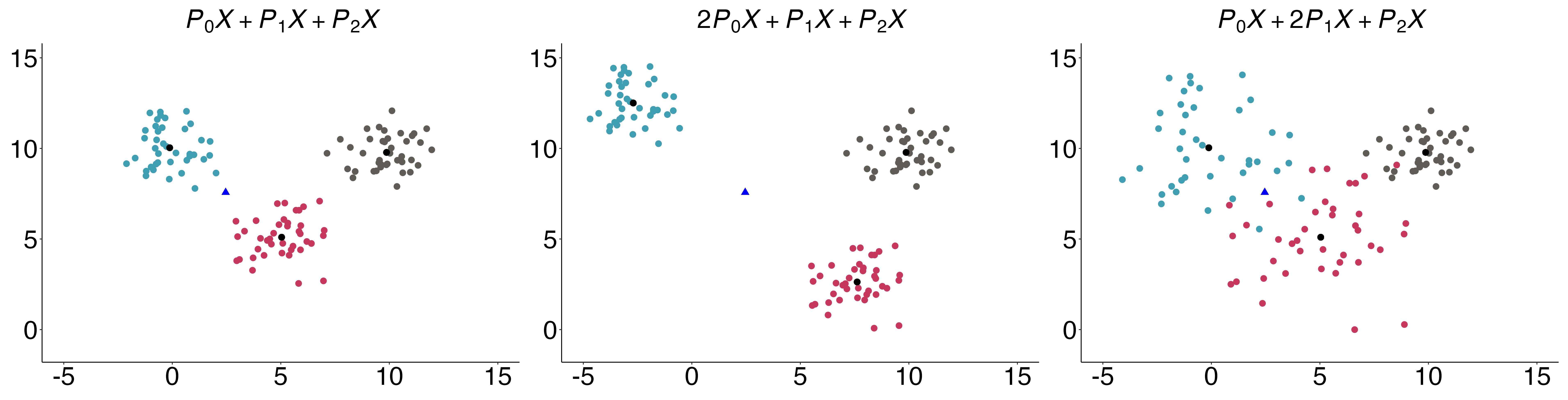}}}
    \caption{Left: visualization of a dataset with colors indicating the clusters formed by the clustering algorithm with $C_k$ represented in blue and $C_{k'}$ in red. The blue triangle represents the combined mean of $C_k\cup C_{k'}$, and the black dots represent the cluster means. Middle: the original dataset with $\|\mathcal{P}_0X\|_F$ scaled by a factor of 2, which pushes apart the clusters $C_k$ and $C_{k'}$. Right: the original dataset with $\|\mathcal{P}_1X\|_F$ scaled by a factor of 2, which spreads points in $C_k$ apart from each other while preserving the cluster mean, and same for points in $C_{k'}$.}\label{fig_decomposition}
\end{figure}

To define our test statistic, we begin by taking a decomposition of the observed data $X$. This decomposition
plays an analogous role to the decomposition~\eqref{eqn:decomp_Gao} used by~\citet{gao2020selective}, but 
is more complex to allow us to handle unknown variance. 
We begin by writing
\[X = \mathcal{P}_0 X + \mathcal{P}_1X + \mathcal{P}_2 X,\]
where $\mathcal{P}_0=\frac{vv^\top}{\|v\|_2^2}$ is the rank-one projection matrix that captures the difference in cluster means for $C_k$ and $C_{k'}$, while
\[\mathcal{P}_1= \left(\mathbf{I}_{C_k} - \frac{\mathbf{1}_{C_k}\mathbf{1}_{C_k}^\top}{|C_k|}\right)  + \left(\mathbf{I}_{C_{k'}} - \frac{\mathbf{1}_{C_{k'}}\mathbf{1}_{C_{k'}}^\top}{|C_{k'}|}\right) ,\]
where, for any subset $C\subseteq[n]$, $\mathbf{I}_C$ represents the diagonal matrix with 
entry $(i,i)$ set to $1$ if $i\in C$ and $0$ if $i\not \in C$. Finally,
\[\mathcal{P}_2 =\mathbf{I}_n-\mathcal{P}_0 - \mathcal{P}_1\]
is the projection matrix to the orthogonal complement of $\mathcal{P}_0$ and $\mathcal{P}_1$. 
We can see that $\mathcal{P}_0$, $\mathcal{P}_1$, and $\mathcal{P}_2$ project to subspaces of dimension $1$, $m-2$, and $n-m+1$, respectively,
where $m = |C_k| + |C_{k'}|$ is the number of data points in the two clusters.
Intuitively, we can think of this decomposition of the data as follows:
\begin{itemize}
\item $\mathcal{P}_0X$ captures the difference in means between clusters $C_k$ and $C_{k'}$;
\item $\mathcal{P}_1X$ captures differences among points within $C_k$, and among points within $C_{k'}$;
\item $\mathcal{P}_2X$ captures all other aspects of the data (i.e., the overall mean of the combined clusters $C_k\cup C_{k'}$, as well
as information about data points not lying in $C_k\cup C_{k'}$). 
\end{itemize}
Figure \ref{fig_decomposition} illustrates the roles of these three terms in the decomposition of the data $X$.

\subsection{The test statistic}

In \citet{gao2020selective}'s work, the test statistic they use is  equivalent to
$\|\mathcal{P}_0 X\|_F$, which under the null hypothesis $H_0$, follows a $\chi_q$ distribution (rescaled by $\sigma$), truncated
to a region $\mathcal{S}$ that controls for the selection event.
In our work, since $\sigma$ is unknown, we will use an F distribution in place of the $\chi$. 
The test statistic we propose is  given by the ratio
\[R = (m-2) \frac{\|\mathcal{P}_0X\|^2_F}{\|\mathcal{P}_1X\|_F^2}.\]
where the numerator is the same (up to squaring and rescaling) as the statistic used by  \citet{gao2020selective}, while the denominator
acts as an estimate of the noise variance $\sigma^2$ (up to rescaling).
We can gain additional intuition for this test statistic by considering the case $K=2$---if we define $T_{\textnormal{two-sample}}$
as the test statistic for the standard two-sample t-test (i.e., testing for equality of means in clusters $k,k'$, assuming equal variances),
we can observe that $R = T_{\textnormal{two-sample}}^2$ in this case.

We next need to define the truncation set. First, we rewrite our decomposition as
\begin{equation}\label{eqn:decomp}X = \|\mathcal{P}_0 X\|_F \cdot \frac{\mathcal{P}_0 X}{\|\mathcal{P}_0 X\|_F}  + \|\mathcal{P}_1X\|_F \cdot \frac{\mathcal{P}_1 X}{\|\mathcal{P}_1 X\|_F} + \mathcal{P}_2 X.\end{equation}
Our test will condition on:
\begin{itemize}
\item The total squared norm $\|\mathcal{P}_0 X\|^2_F +  \|\mathcal{P}_1 X\|^2_F$ for the first and second terms in the decomposition;
\item The normalized terms $ \frac{\mathcal{P}_0 X}{\|\mathcal{P}_0 X\|_F}$ and $\frac{\mathcal{P}_1 X}{\|\mathcal{P}_1 X\|_F} $
for the first and second terms in the decomposition;
\item The third term $\mathcal{P}_2 X$ in the decomposition.
\end{itemize}
With these terms treated as known, the data $X$ can then be fully determined
by revealing the value $R =(m-2) \frac{\|\mathcal{P}_0X\|^2_F}{\|\mathcal{P}_1X\|_F^2}$ of the test statistic. For any $ r>0$, define
\[x'( r ) 
= \left(\sqrt{ \frac{r}{m-2+r} } \cdot \frac{\mathcal{P}_0 X}{\|\mathcal{P}_0 X\|_F}+\sqrt{\frac{m-2}{m-2+r} }\cdot \frac{\mathcal{P}_1 X}{\|\mathcal{P}_1 X\|_F}  \right)  \cdot \sqrt{ \|\mathcal{P}_0 X\|^2_F +  \|\mathcal{P}_1 X\|^2_F }+ \mathcal{P}_2X.
\]
We can verify from the definition of $R$ that
$X = x'(R)$ holds by definition.
Finally, define
\[\mathcal{S}' = \left\{ r>0 : \textnormal{Cluster}(X)=\textnormal{Cluster}(x'( r ))\right\}\subseteq(0,\infty).\]

\subsection{Main result}

Our main result, presented next, establishes that we can compute the exact post-selection distribution of $R$,
which thus allows us to perform valid selective inference in the unknown-variance setting.
\begin{theorem}\label{thm:main}
Let $X_i\indsim \mathcal{N}(\mu_i,\sigma^2\mathbf{I}_q)$ where $\sigma$ is unknown, and let $\mathcal{P}_0$, $\mathcal{P}_1$, and $\mathcal{P}_2$ be defined as above. Then, conditional on $\textnormal{Cluster}(X)$, $\|\mathcal{P}_0X\|_F^2+\|\mathcal{P}_1X\|_F^2$, $\frac{\mathcal{P}_0X}{\|\mathcal{P}_0X\|_F}$, $\frac{\mathcal{P}_1X}{\|\mathcal{P}_1X\|_F}$, and $\mathcal{P}_2X$, under the null hypothesis $H_0'\left(C_k,C_{k'}\right)$, the random variable $R$ follows the 
$\textnormal{F}_{q,(m-2)q}$ distribution
truncated to the set $\mathcal{S}'$. In particular, the p-value
\[P'=1-F_{\textnormal{F}_{q,(m-2)q}}\left(R;\mathcal{S}'\right)\]
is uniformly distributed under $H_0'\left(C_k,C_{k'}\right)$, where $F_{\textnormal{F}_{q,(m-2)q}}(\cdot;\mathcal{S}')$ is the CDF of 
a $\textnormal{F}_{q,(m-2)q}$ random variable truncated to the set $\mathcal{S}'$.
\end{theorem}
The intuition is that, if $\mathcal{P}_0$ and $\mathcal{P}_1$ were fixed rather than data-dependent (i.e., if the clusters $C_k$ and $C_{k'}$
were chosen before viewing the data), then we would have $\sigma^{-2}\|\mathcal{P}_0 X\|^2_F\sim \chi^2_q$ and, independently, $\sigma^{-2}\|\mathcal{P}_1X\|^2_F\sim\chi^2_{(m-2)q}$;
 thus $R =  (m-2)\frac{\|\mathcal{P}_0X\|^2_F}{\|\mathcal{P}_1X\|_F^2}$ would follow an $F_{q,(m-2)q}$ distribution. After accounting
for the selection event, the null distribution is instead given by a truncated F distribution.

To implement the results of Theorem \ref{thm:main} in practice, we need to be able to compute this p-value. In other words, we need to either  explicitly characterize the set $\mathcal{S}'$  that is consistent with the selection event, or develop an empirical sampling strategy to estimate the p-value. We next consider this computational question.

\subsection{Computing the p-value}\label{subsec:charac}
To characterize the truncation set $\mathcal{S}'$, we will split into two cases. In the general setting, when the data is separated into an arbitrary number $K\geq 2$ 
of clusters, 
we will handle the truncation event via numerical approximation. For the special case $K=2$,
however, we will show that $\mathcal{S}'$ can potentially be computed explicitly, by relating the problem back to the work of \citet{gao2020selective} for the 
known-variance case.

\subsubsection{Special case: $K= 2$}\label{sec:computation_exact}
Rewriting \citet{gao2020selective}'s procedure in our notation, the modified data is defined as
\begin{equation}\label{eqn:x(phi)_in_our_notation}x(\phi) = \frac{\phi}{\|v\|_2}  \cdot \frac{\mathcal{P}_0 X}{\|\mathcal{P}_0 X\|_F} + \mathcal{P}_1 X + \mathcal{P}_2 X,\end{equation}
where $v = \frac{\mathbf{1}_{C_k}}{|C_k|} -  \frac{\mathbf{1}_{C_{k'}}}{|C_{k'}|}$ so that $\mathcal{P}_0 = \frac{vv^\top}{\|v\|^2_2}$
is projection onto the span of $v$,
and their selection set is given by $\mathcal{S} = \{\phi > 0 :  \textnormal{Cluster}(X)=\textnormal{Cluster}(x(\phi))\}$.
In contrast, our method defines
\[x'( r ) 
= \left(\sqrt{ \frac{r}{m-2+r} } \cdot \frac{\mathcal{P}_0 X}{\|\mathcal{P}_0 X\|_F}+\sqrt{\frac{m-2}{m-2+r} }\cdot \frac{\mathcal{P}_1 X}{\|\mathcal{P}_1 X\|_F}  \right) \cdot \sqrt{ \|\mathcal{P}_0 X\|^2_F +  \|\mathcal{P}_1 X\|^2_F }+ \mathcal{P}_2X
\]and $\mathcal{S}' = \{ r>0:   \textnormal{Cluster}(X)=\textnormal{Cluster}(x'( r ))\}$.

The next result shows that, for the case $K=2$, these two definitions can be related in a simple way.
\begin{proposition}\label{prop:trunc_relate}
Suppose that $K=2$ (so that we have $k=1$, $k'=2$). Assume
that  the clustering procedure is location- and scale-invariant---that is, for any $x\in\R^{n\times q}$,
\[\textnormal{Cluster}(x) = \textnormal{Cluster}\left(a \cdot x + \mathbf{1}_n \cdot b^\top\right),\]
for any $a\in\R$ and $b\in\R^q$. Then, under the notation and definitions above,
\[\mathcal{S}' = \left\{ r>0 :\sqrt{ \frac{ r }{m-2 } }\cdot \|\mathcal{P}_1 X\|_F \cdot \sqrt{\frac{1}{|C_1|} + \frac{1}{|C_2|}} \in\mathcal{S}\right\}.\]
\end{proposition}
\noindent The work of  \citet{gao2020selective} gives an explicit characterization of the set $\mathcal{S}$
for a family of agglomerative hierarchical clustering algorithms,  while \cite{chen2022selective} does the same for k-means clustering.
Moreover, both of these algorithms are location- and scale-invariant.
Consequently, for the case $K=2$, we can explicitly characterize the truncation set $\mathcal{S}'$, and thus can compute the p-value $P'$ constructed 
in Theorem~\ref{thm:main} by leveraging \citet{gao2020selective}'s construction of the set $\mathcal{S}$, for these two popular algorithms.

Of course, a major limitation of this result is that we can only handle $K=2$. However, in iterative procedures
(e.g., hierarchical clustering), the test can be applied at the first step (i.e., the first split, from a single cluster to $K=2$ clusters).
This hypothesis
test can then essentially be interpreted as testing the global null, i.e., whether the data should be split into clusters at all
or simply treated as a single cluster.

\subsubsection{General case: $K\geq 2$}\label{sec:computation_general}

Next we consider the general case.
If $K>2$, then the result of Proposition~\ref{prop:trunc_relate} no longer applies---as we will see in the proof of this proposition, the case $K=2$
leads to the result specifically because $\mathcal{P}_2 = \frac{\mathbf{1}_n\mathbf{1}_n^\top}{n}$ in that case, but this no longer holds for $K>2$.
Moreover, even if $K=2$, it might be the case that 
 we are using a clustering algorithm for which $\mathcal{S}$ does not have an explicit characterization and/or
 which is not location- and scale-invariant. In any of these settings, 
 we do not have an explicit characterization of $\mathcal{S}'$, and thus the truncated CDF
 needed in Theorem~\ref{thm:main} cannot be computed exactly.
 
 Nonetheless, the inference procedure can still be run in the general case, by computing $P'$ approximately through importance sampling. 
 Specifically, in order to (approximately) compute the p-value $P'$, we need to be able to estimate
 \begin{multline*}1-F_{\textnormal{F}_{q,(m-2)q}}\left(r;\mathcal{S}'\right)
  = \Ppst{R'\sim \textnormal{F}_{q,(m-2)q}}{R'>r}{R'\in\mathcal{S}'}\\
  = \frac{\Pp{R'\sim \textnormal{F}_{q,(m-2)q}}{R'>r, \ R'\in\mathcal{S}'}}{\Pp{R'\sim \textnormal{F}_{q,(m-2)q}}{R'\in\mathcal{S}'}}, \end{multline*}
and then evaluate this probability at $r =R$.
Since $r=R$ may be extremely large, and/or the selection set $\mathcal{S}'$ may be very small,
the events $R'>r$ and/or $R'\in\mathcal{S}'$ may 
have extremely low probability under
the distribution $R'\sim\textnormal{F}_{q,(m-2)q}$. We approximate this probability with importance sampling,
using a truncated normal distribution as the proposal distribution. Further details are given in Appendix~\ref{app:importance_sampling}.

\section{Empirical results}\label{sec:empirical}

We now provide empirical results for our proposed method. We present simulation results for Type I error control and empirical power for our proposed
method, as well as results on a small real dataset (the penguin dataset provided by \citep{horst2020palmerpenguins}, which was also analyzed in \citet{gao2020selective}).
For all experiments, we use the hierarchical clustering algorithm with average linkage for clustering.\footnote{Code for reproducing all experiments is available online at \url{https://github.com/yjyun97/cluster_inf_unknown_var}.}

\subsection{Type I error control}\label{sec:empirical_null}
Theorem~\ref{thm:main} states that $P'$ follows the uniform distribution under the null, so it controls the Type 1 error rate of the hypothesis test. We illustrate this empirically by plotting empirical quantiles of a sample of $P'$ against the theoretical quantiles of the uniform distribution. We compare to the p-value $P$ computed by  \citet{gao2020selective}'s method, with either oracle knowledge of the true $\sigma$,
or with plug-in estimates  $\hat\sigma_{\textnormal{all}}$ or $\hat\sigma_{\textnormal{clustered}}$.
(The results for the latter two variants of  \citet{gao2020selective}'s method were also shown in the top row of Figure~\ref{fig_motivation} in Section~\ref{sec:background_challenges}.)

\begin{figure}[hbtp]
    \centering
    {{\includegraphics[width=15cm]{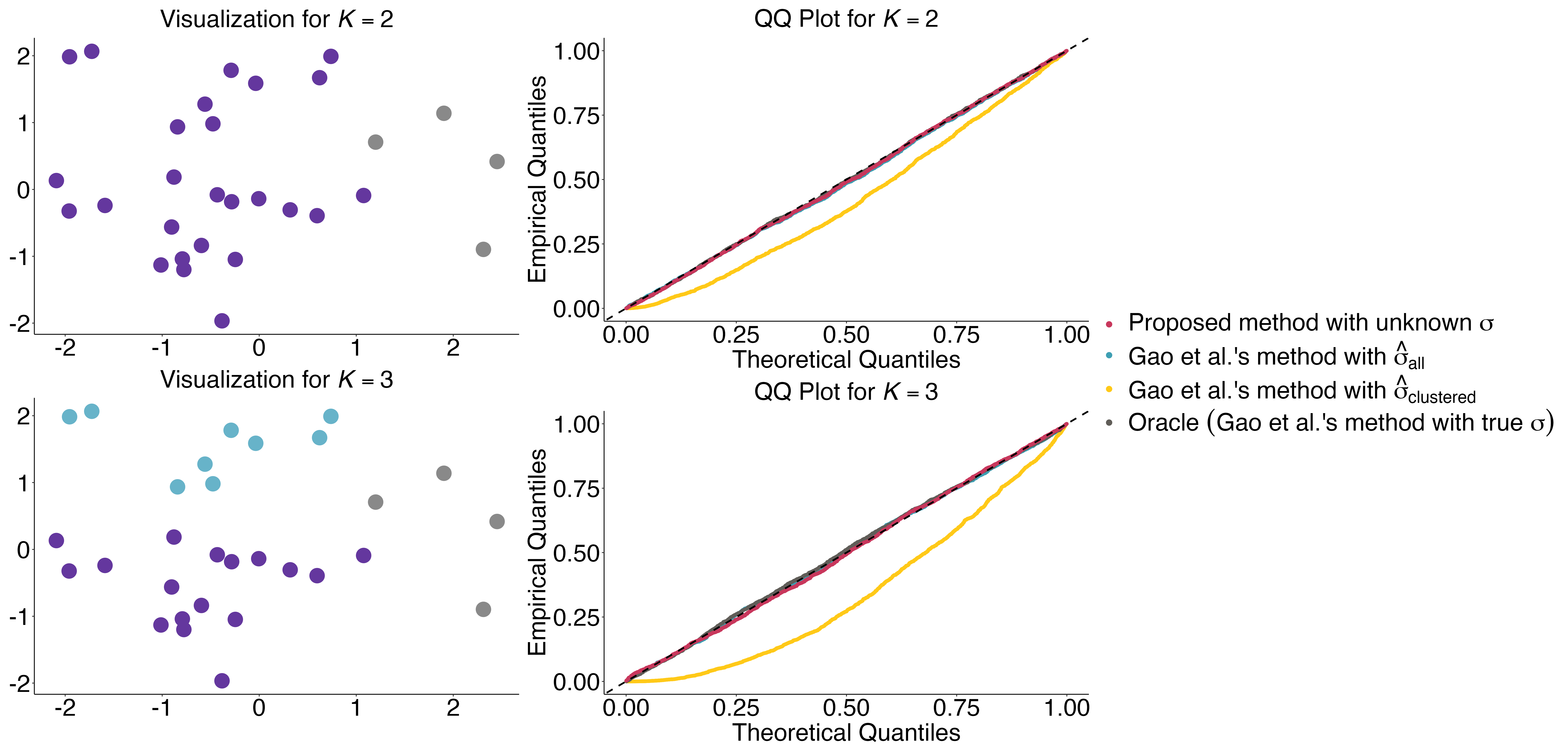}}}
    \caption{Simulation under the null (see Section~\ref{sec:empirical_null} for details). The left plots show one draw of the data, along with the
    result of hierarchical clustering for $K=2$ (top) and $K=3$ (bottom); in these figures,
    and in analogous figures below, the shape of each data point
     indicates the true cluster from which the data point originated, while the color indicates the cluster assignment estimated from the data.
    The right plots show QQ plots comparing the p-values obtained via the four different methods.
}\label{fig_type1}
\end{figure}

We perform 2000 independent trials, with datasets generated as  \[X_i\indsim \mathcal{N}(\mu_i,\sigma^2\mathbf{I}_q)\]
for $i\in[n]$, with $\mu_i=0_q$ for all $i\in[n]$ so that the null hypothesis is true. We set $\sigma=1$, $n=30$, and $q=2$. Figure~\ref{fig_type1} presents results for two different settings,
$K=2$ and $K=3$. We use the exact computation method presented in Section \ref{sec:computation_exact} for $K=2$, and the importance sampling method discussed in \ref{sec:computation_general} for $K=3$ (with $N=8000$ draws when we run importance sampling). 
For the $K=2$ case, the p-values are generated for the comparison between clusters $k=1$ and $k'=2$.
For the $K=3$ case, in each trial, the p-values are generated for the comparison between two randomly chosen clusters $k\neq k'\in\{1,2,3\}$.

Figure~\ref{fig_type1} illustrates that the proposed method,
along with \citet{gao2020selective}'s method with either the conservative estimate $\hat\sigma_{\textnormal{all}}$ or with oracle knowledge of the true $\sigma$, all result in uniformly
distributed p-values (and thus, control the Type I error rate) for both settings.
On the other hand, \citet{gao2020selective}'s method applied with $\hat\sigma_{\textnormal{clustered}}$  fails to do so, with a substantially anticonservative 
distribution of the p-values. As expected, since $\hat\sigma_{\textnormal{clustered}}$ is a more extreme underestimate for larger $K$,
the nonuniformity of the p-values is more substantial when $K=3$ than when $K=2$.

\subsection{Empirical power}\label{sec:empirical_power}
It is inevitable that not knowing $\sigma$ will cause some loss in power, as we are forced to condition on more components of $X$ than we would otherwise do when computing the p-value. On the other hand, using $\hat{\sigma}_{\textnormal{all}}$ in place of $\sigma$ in $P$ also causes a loss in power, due to $\hat{\sigma}_{\textnormal{all}}$ being an overestimate of $\sigma$ in settings where the alternative hypothesis is true. Here, we compute the empirical power of the same four methods as above.
To be more precise, to calculate empirical power, for each trial, we again consider clusters $k=1$ and $k'=2$ for the case $K=2$, or a randomly chosen pair $k\neq k'\in\{1,2,3\}$ for the case $K=3$.
For each trial, we first verify whether the null $H_0'(C_k,C_{k'})$ holds---in other words, if the recovered clusters $C_k,C_{k'}$ are both subsets of the \emph{same}
original true cluster. Empirical power is then defined as the proportion of times that we reject the null, among all trials for which this null is false.

\begin{figure}[t]
    \centering
    {{\includegraphics[width=15cm]{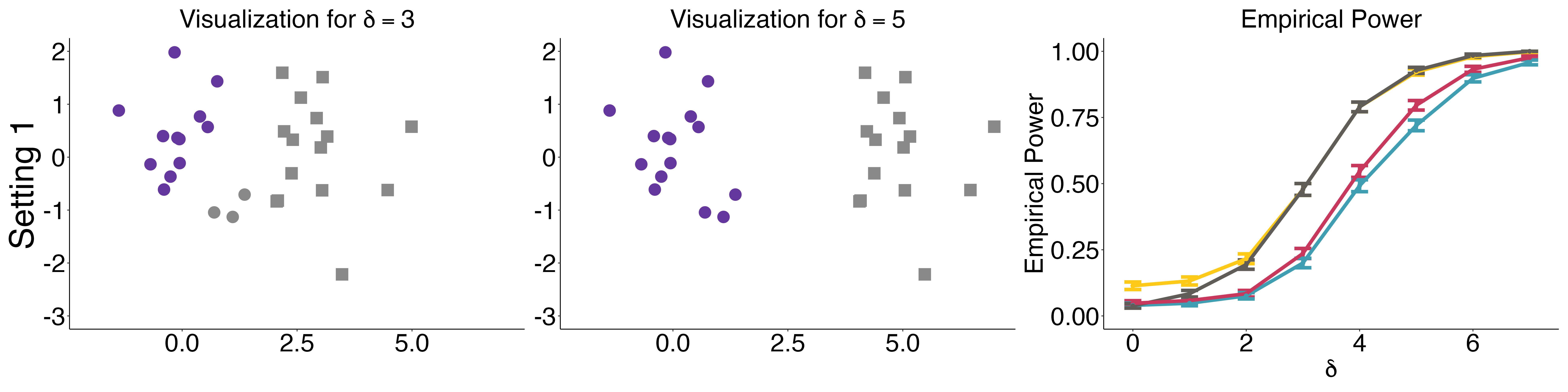}}}\\
        {{\includegraphics[width=15cm]{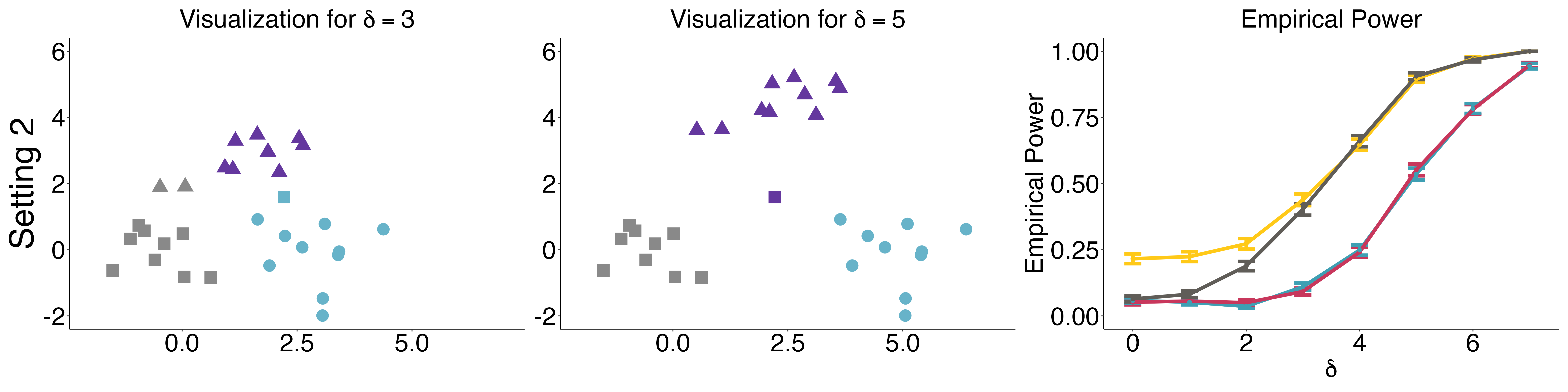}}}\\
            {{\includegraphics[width=15cm]{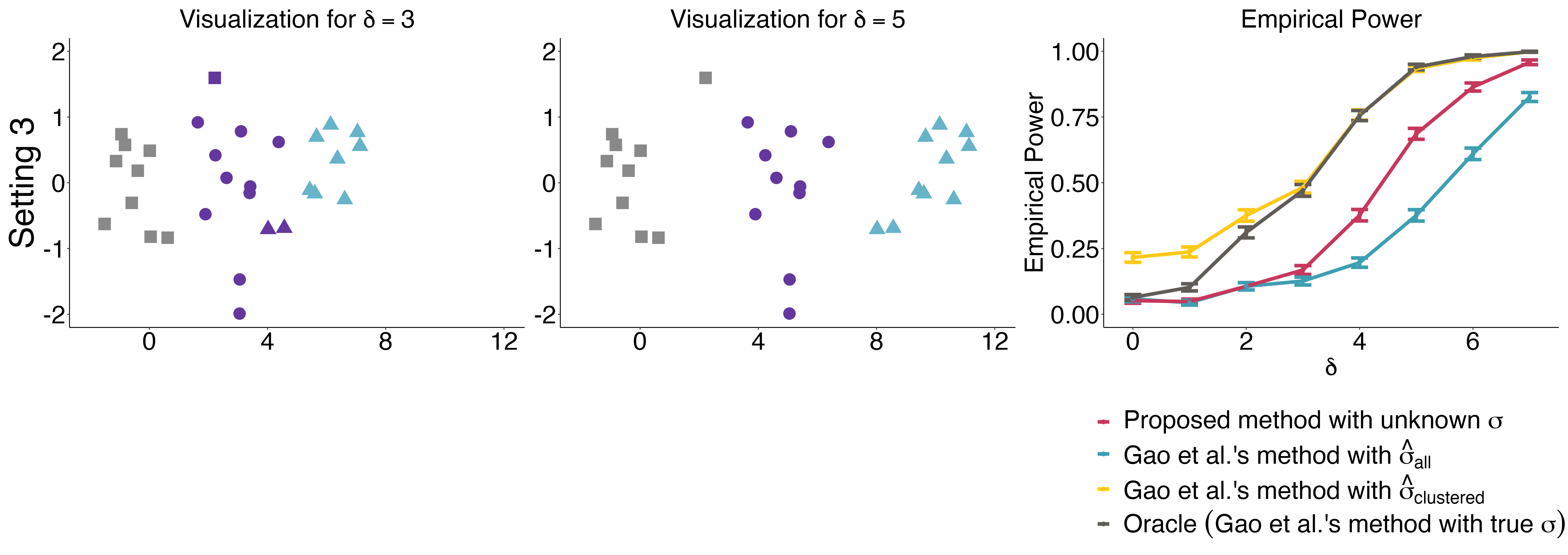}}}
    \caption{Simulation under the alternative, for Setting 1 (top row), Setting 2 (middle row), and Setting 3 (bottom row). (See Section~\ref{sec:empirical_power} for details.) The left and middle column of plots show one draw of the data for two different signal strengths $\delta$, along with the
    result of hierarchical clustering. 
    The right column of plots shows power as a function of signal strength $\delta$ for the four different methods.
    }\label{fig_power} 
\end{figure}

We perform 500 independent trials, with datasets generated as 
\[X_i\indsim \mathcal{N}(\mu_i,\sigma^2\mathbf{I}_q)\] for $i\in [n]$ with $\sigma=1$, $n=30$, and $q=2$.
As before, we use the exact computation method presented in Section \ref{sec:computation_exact} for $K=2$, and the importance sampling method discussed in \ref{sec:computation_general} for $K=3$ (with $N=8000$ draws when we run importance sampling).
We test three different settings for generating the means $\mu_i$:
\begin{itemize}
\item Setting 1: $K=2$ true clusters,  with $\mu_i=(0,0)^\top$ for $n/2=15$ data points, and $\mu_i=(\delta,0)^\top$ for the remaining $n/2=15$ data points.
(The results for the three variants of  \citet{gao2020selective}'s method in Setting 1 were also shown in the bottom row of Figure~\ref{fig_motivation} in Section~\ref{sec:background_challenges}.)
\item Setting 2: $K=3$ true clusters,  with $\mu_i=(0,0)^\top$ for $n/3=10$ data points,  $\mu_i=(\delta,0)^\top$ for $n/3=10$ data points, and $\mu_i=(\delta/2,\delta\sqrt{3}/2)^\top$ for the remaining $n/3=10$ data points.
\item Setting 3: $K=3$ true clusters,  with $\mu_i=(0,0)^\top$ for $n/3=10$ data points,  $\mu_i=(\delta,0)^\top$ for $n/3=10$ data points, and $\mu_i=(2\delta,0)^\top$ for the remaining $n/3=10$ data points.
\end{itemize}
We then run hierarchical clustering with the true number of clusters $K$.
In each setting, the parameter $\delta \in\{0,1,\dots,7\}$ controls the signal strength.
Note that $\delta=0$ corresponds to the null(s) being true, as all data points have mean $\mu_i=0$. Setting 1 with $\delta=0$ is identical to the first simulation under the null in
Section~\ref{sec:empirical_null}  (where hierarchical clustering is run with $K=2$), while Settings 2 and 3 with $\delta=0$ are identical to the second simulation under the null in
Section~\ref{sec:empirical_null} (where hierarchical clustering is run with $K=3$).
 For the $K=2$ case (Setting 1), the p-values are generated for the comparison between clusters $k=1$ and $k'=2$.
For the $K=3$ case (Settings 2 and 3), in each trial, the p-values are generated for the comparison between two randomly chosen clusters $k\neq k'\in\{1,2,3\}$.
In all settings, the threshold for rejecting a p-value is $\alpha = 0.05$.

Figure~\ref{fig_power}  illustrates the power, as a function of $\delta$, for the four methods in each of the three settings.
We see that the highest power is achieved by \citet{gao2020selective}'s method applied with the anticonservative estimate $\hat\sigma_{\textnormal{clustered}}$, 
but this power comes at the cost of loss of Type I error control (as we can see due to the high rejection rate at $\delta=0$, where the null is true).
Among the remaining methods, we see that the proposed method always has power at least as high as 
\citet{gao2020selective}'s method applied with the conservative estimate $\hat\sigma_{\textnormal{all}}$---sometimes approximately
the same, and sometimes substantially higher, in the various settings. Determining the types of settings where our proposed method will result in a substantial gain in power remains an interesting open question.

\subsection{Robustness to model misspecification}
In this section, we empirically study the robustness of the proposed method to model misspecification for two different settings, with comparison to that of \citet{gao2020selective}'s method applied with $\hat\sigma_{\textnormal{all}}$ and $\hat\sigma_{\textnormal{clustered}}$. 

\begin{figure}[hbtp]
    \centering
    {{\includegraphics[width=15cm]{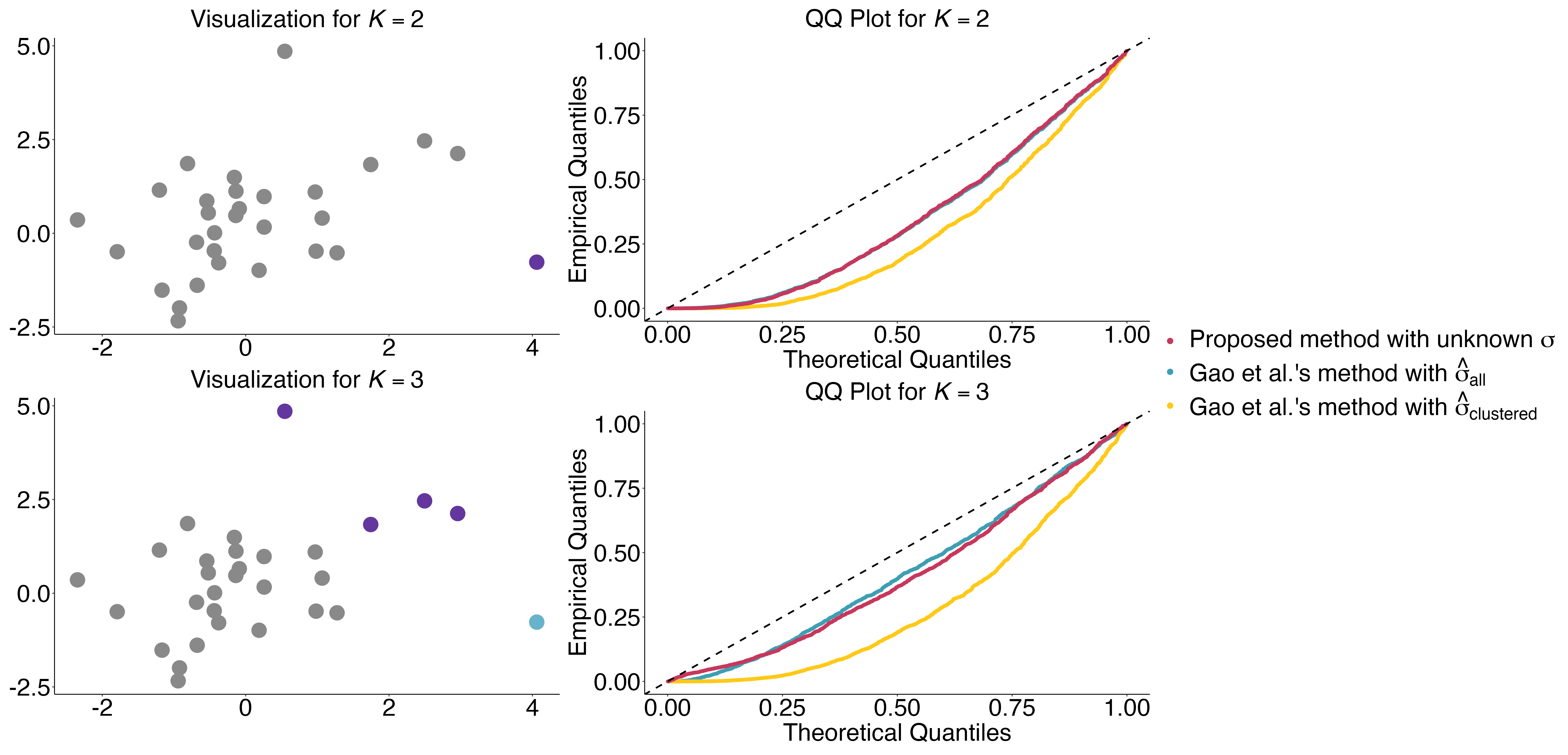}}}
    \caption{Simulation under the null under model misspecification, with noise generated from an extremely heavy-tailed distribution, $t_5$.
}\label{fig:heavy5}
\end{figure}

\begin{figure}[hbtp]
    \centering
    {{\includegraphics[width=15cm]{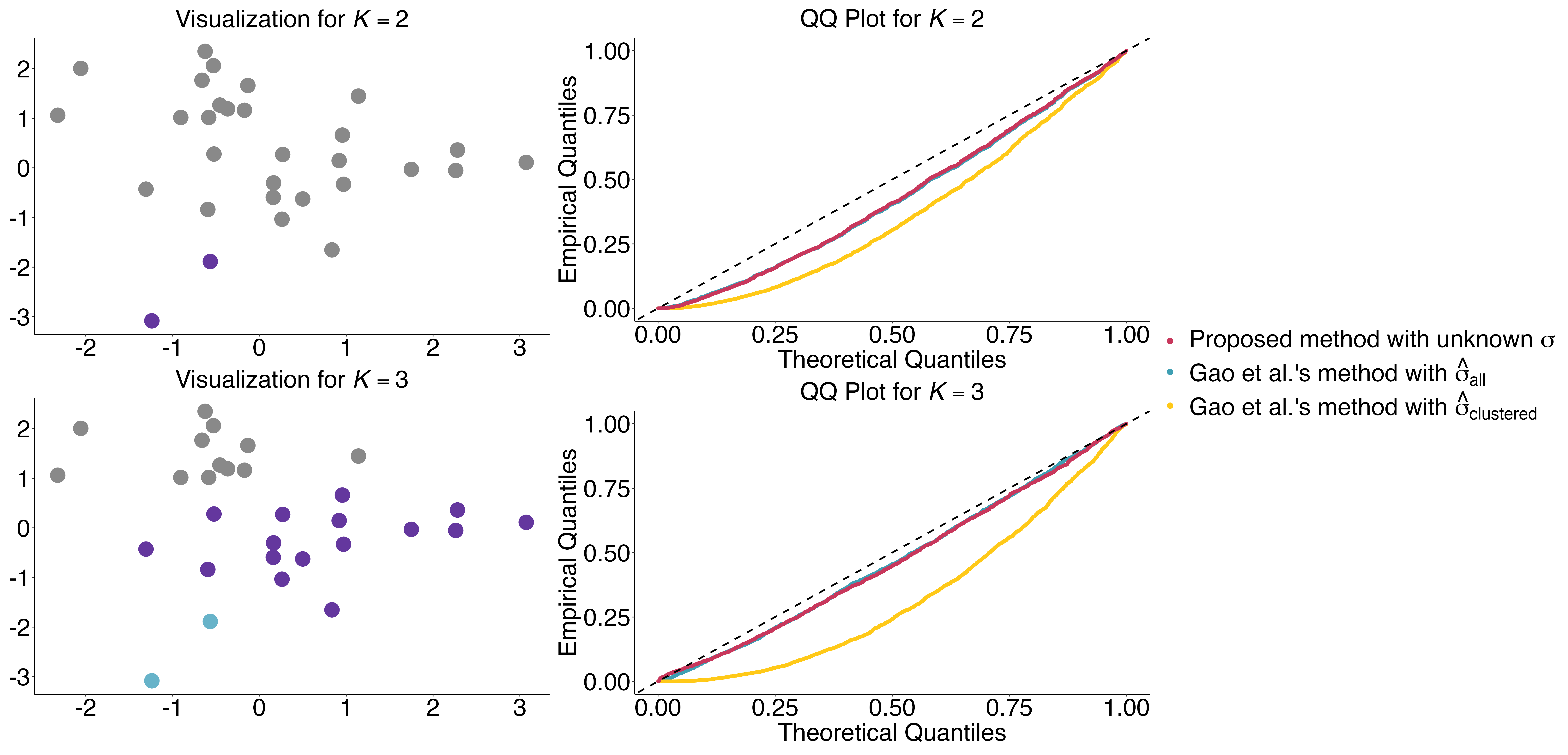}}}
    \caption{Simulation under the null under model misspecification, with noise generated from a moderately heavy-tailed distribution, $t_{10}$.}
\label{fig:heavy10}
\end{figure}

\begin{figure}[hbtp]
    \centering
    {{\includegraphics[width=15cm]{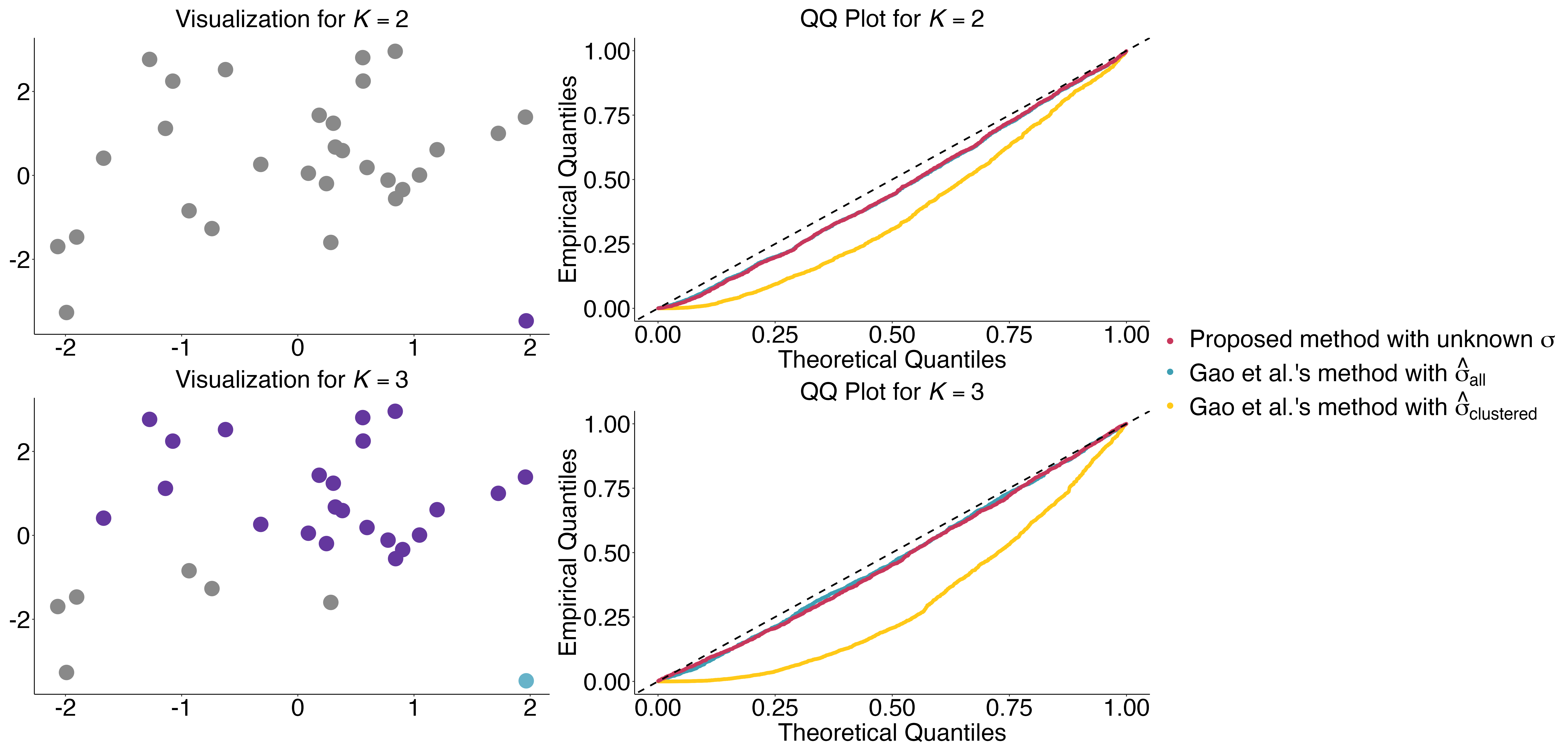}}}
    \caption{Simulation under the null under model misspecification, with noise generated from a Gaussian distribution with a non-isotropic covariance matrix.
}\label{fig:non-iso}
\end{figure}

We reproduce the experiments in Section \ref{sec:empirical_null} (where $\mu_i\equiv 0$, i.e., all data is drawn from the null), with the only difference being the data generating process. 
First we consider a setting where the data are generated from an extremely heavy-tailed distribution, 
\[X_{ij}\iidsim t_5, \ i\in[n],\ j\in [q].\]
We next repeat with a more moderately heavy-tailed distribution,
\[X_{ij}\iidsim t_{10}, \ i\in[n],\ j\in [q].\]
Finally, we consider a Gaussian distribution with a non-isotropic covariance matrix. 
\[X_i\iidsim \mathcal{N}\left(\left(\begin{array}{c}0\\0\end{array}\right),\left(\begin{array}{cc} 1 & 0 \\ 0 & 2\end{array}\right)\right), \ i\in [n].\]
The results for these three settings are shown in
Figures~\ref{fig:heavy5}, \ref{fig:heavy10}, and \ref{fig:non-iso}, respectively.
 In all three settings, we see that all of the methods show inflation of Type I error in the presence of model misspecification.
As expected, this is more severe for the extremely heavy tailed $t_5$ noise, and more mild for the other two settings. Moreover,
as expected, \citet{gao2020selective}'s method with the anti-conservative estimate $\hat\sigma^2_{\textnormal{clustered}}$ shows substantially more 
inflation of the Type I error; the inflation is similar between our method and \citet{gao2020selective}'s method with the anti-conservative estimate $\hat\sigma^2_{\textnormal{all}}$.

\subsection{Real dataset}
\begin{figure}[t]
    \centering
    {{\includegraphics[width=10cm]{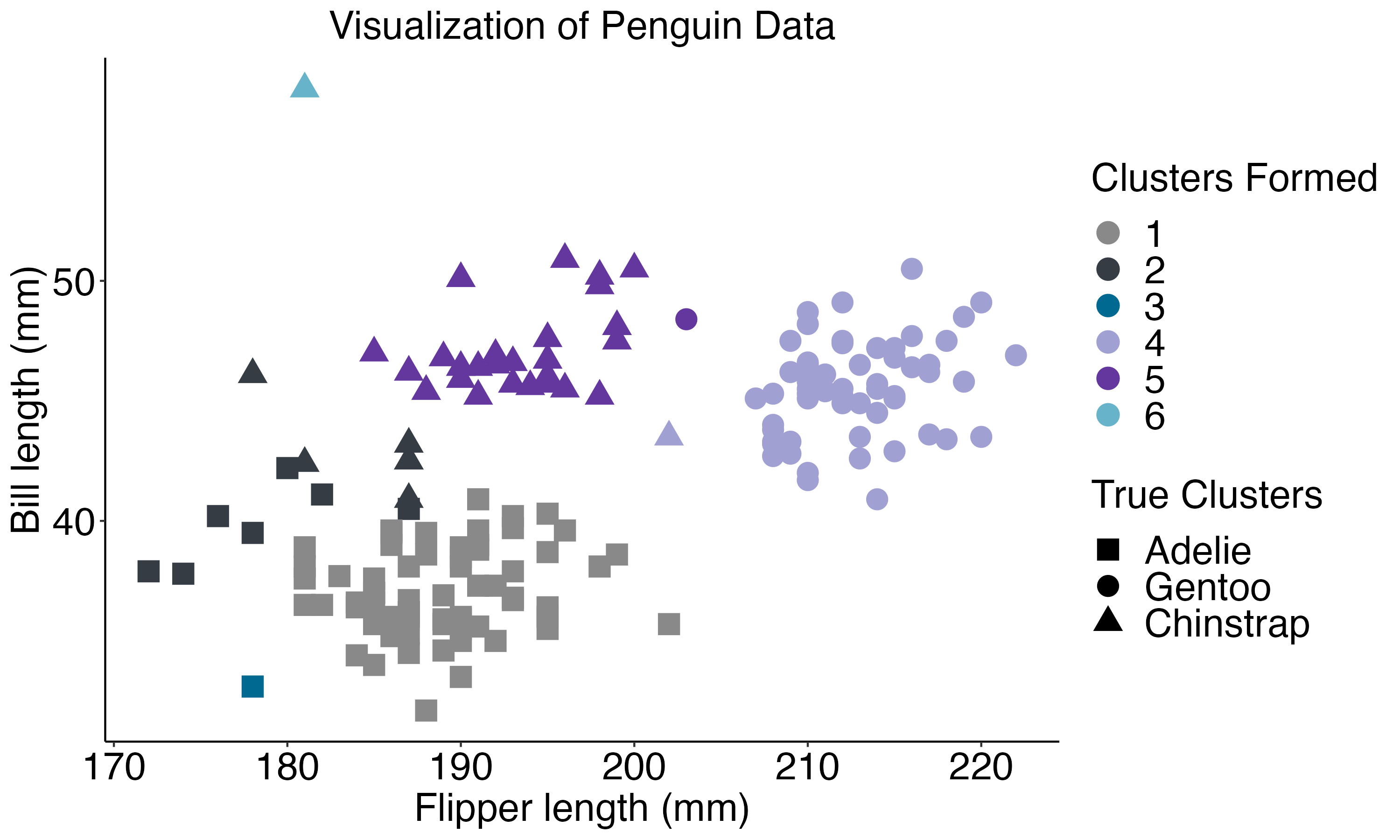}}}
    \caption{Visualization of the penguin dataset. The output of the clustering algorithm corresponds to that run on the centered and standardized dataset.}\label{fig_penguin}
\end{figure}

We next compare the  methods on a real dataset---the penguin dataset \citep{horst2020palmerpenguins}, which contains information about the bill length (mm) and flipper length (mm) of three different species of penguins, Adelie, Chinstrap, and Gentoo. (This dataset was also studied by \citet[Section 6]{gao2020selective} to test their method for inference after
clustering.) The data is shown in Figure \ref{fig_penguin}.

 Table~\ref{table_penguin} shows the p-values computed by each of the three methods that does not require knowledge
 of $\sigma$---our proposed method, along with \citet{gao2020selective}'s with $\hat\sigma_{\textnormal{all}}$ or with $\hat\sigma_{\textnormal{clustered}}$.
 We test three different pairwise cluster comparisons (i.e., comparing $C_k$
and $C_{k'}$, for three different choices of $(k,k')$).

\begin{table}[t]
\centering
\begin{tabular}{rrrr}
  \hline
 $(k,k')$& $(1,2)$ & $(1,5)$ & $(4,5)$ \\ 
  \hline
Proposed method & 0.5 & 0.0045 & 1.5e-08 \\ 
  \citet{gao2020selective}'s method with $\hat\sigma_{\textnormal{all}}$ & 0.85 & 0.13 & 0.0014 \\ 
  \citet{gao2020selective}'s method with $\hat\sigma_{\textnormal{clustered}}$ & 0.31 & 1.6e-07 & 4.2e-22 \\ 
   \hline
\end{tabular}
\caption{Results on the centered and standardized penguin dataset.}\label{table_penguin}
\end{table}

Since the true clustering (i.e., the penguin species) is known, we can see that $(k,k') = (1,2)$ corresponds to two estimated clusters that
are not very well separated in terms of the true species labeling,
while $(k,k')=(1,5)$ and $(k,k') = (4,5)$ clearly correspond to a true difference in species.
In Table~\ref{table_penguin}, overall we see that
the proposed method produces p-values that are substantially lower than those computed by   \citet{gao2020selective}'s method with
the conservative estimate $\hat\sigma_{\textnormal{all}}$, corresponding to a gain in power.
  \citet{gao2020selective}'s method applied with $\hat\sigma_{\textnormal{clustered}}$ produces p-values even lower than our proposed
  method, but may not be reliable in terms of Type I error control as we have seen in our simulations.

\section{Discussion}\label{sec:summary}
In this work, we have proposed an extension of \citet{gao2020selective}'s framework for selective inference for clustering to the case where the isotropic covariance matrix is unknown. 
Since the method does not rely on plug-in empirical estimates of $\sigma$, we can avoid loss of power and/or loss of Type I error control.
For location- and scale-invariant clustering algorithms, we have shown that the resulting p-value can be computed exactly in the case of $K=2$ clusters by leveraging connections
to the findings of \citet{gao2020selective}; more generally, standard sampling strategies allow for accurate estimation of the p-value.

These results suggest a number of possible extensions and open questions. First, in our work we allow for unknown variance but assume an isotropic covariance structure, i.e., $\sigma^2\mathbf{I}_q$, for the $q$-dimensional data points. It would be interesting to extend these techniques to the setting of diagonal covariance, or even an arbitrary covariance matrix, to allow for nonconstant variance along the $q$ coordinates and/or nonzero correlation. Another important question is whether these tools can be extended to the non-Gaussian setting, which would offer further flexibility and robustness for real-data applications.

\subsection*{Acknowledgements}
R.F.B. was supported by the National Science Foundation via grants DMS-1654076 and DMS-2023109, and by the Office of Naval Research via grant N00014-20-1-2337.

\bibliographystyle{plainnat}
\bibliography{bib}

\appendix

\section{Additional proofs and details}\label{app:proofs}

\subsection{Proof of Theorem \ref{thm:main}}\label{app:thm_main}

First, suppose that clusters $C_k$ and $C_{k'}$ were fixed, i.e., not data-dependent. We will show that, in that case, we have 
\[R \,\mid\, Z\sim \textnormal{F}_{q,(m-2)q}\]
under $H_0'\left(C_k,C_{k'}\right)$, where
\[Z = \left(\|\mathcal{P}_0X\|_F^2+\|\mathcal{P}_1X\|_F^2,\frac{\mathcal{P}_0X}{\|\mathcal{P}_0X\|_F}, \frac{\mathcal{P}_1X}{\|\mathcal{P}_1X\|_F}, \mathcal{P}_2X\right).\] 
Since $X$ has isotropic covariance while $\mathcal{P}_0,\mathcal{P}_1,\mathcal{P}_2$ are orthogonal, we see that $\mathcal{P}_0 X$, $\mathcal{P}_1 X$, $\mathcal{P}_2 X$ are mutually
independent, and so we now only need to show
\[R \,\mid\,  \left(\|\mathcal{P}_0X\|_F^2+\|\mathcal{P}_1X\|_F^2,\frac{\mathcal{P}_0X}{\|\mathcal{P}_0X\|_F}, \frac{\mathcal{P}_1X}{\|\mathcal{P}_1X\|_F}\right)\sim \textnormal{F}_{q,(m-2)q}.\]
Moreover, 
under $H_0'\left(C_k,C_{k'}\right)$, we see that $\EE{\mathcal{P}_0X} =0$ and $\EE{\mathcal{P}_1 X}=0$; therefore,
by properties of the Gaussian distribution (with mean zero and isotropic covariance), we have $(\|\mathcal{P}_0 X\|_F, \|\mathcal{P}_1X\|_F)$ independent
from $\left(\frac{\mathcal{P}_0X}{\|\mathcal{P}_0X\|_F}, \frac{\mathcal{P}_1X}{\|\mathcal{P}_1X\|_F}\right)$, and so now it suffices to show that 
\[R \,\mid\,  \|\mathcal{P}_0X\|_F^2+\|\mathcal{P}_1X\|_F^2\sim  \textnormal{F}_{q,(m-2)q}.\]
Finally,  $\|\mathcal{P}_0 X\|_F$ and $\|\mathcal{P}_1X\|_F$ are independent, with 
$\sigma^{-2} \|\mathcal{P}_0 X\|^2_F \sim \chi^2_q$ and $\sigma^{-2} \|\mathcal{P}_1X\|^2_F \sim \chi^2_{(m-2)q}$ (because $\mathcal{P}_0$ is a rank-$1$ projection matrix
while $\mathcal{P}_1$ is a rank-$(m-2)$ projection matrix).
The desired statement follows since, for independent random variables $A\sim\chi^2_a$ and $B\sim \chi^2_b$,
it follows from properties of the $\chi^2$ and  F distributions that $\frac{A/a}{B/b}$ is independent from $A+B$, and follows a $\textnormal{F}_{a,b}$
distribution.

Next, we will account for the fact that clusters $C_k$ and $C_{k'}$ are data-dependent,
by conditioning on $\textnormal{Cluster}(X)$.
Let $f_{(R,Z)}( r ,z)$ denote the joint density of $(R,Z)$ with respect to the appropriate base measure, when we treat $\textnormal{Cluster}(X)$ (and thus $C_k,C_{k'}$)  as fixed.
By the work above we can write $f_{(R,Z)}( r ,z) = f_R( r )f_Z(z)$ where $f_R$ is the density of the $\textnormal{F}_{q,(m-2)q}$ distribution. 
We then need to calculate the conditional distribution of $R$, given the event $(R,Z)\in E^*$, where $E^*$ is the subset of all values $( r ,z)$ such that $\textnormal{Cluster}(x'( r )) = \textnormal{Cluster}^*$, for some particular clustering $ \textnormal{Cluster}^*$ (and then we will apply the calculation with $\textnormal{Cluster}^* =  \textnormal{Cluster}(X)$)
(and note that, by the construction of clustering procedures, $(R,Z)\in E^*$ has positive probability).
The density of $(R,Z)\mid (R,Z)\in E^*$ is then given by
\[f_{(R,Z)\mid (R,Z)\in E^*}( r ,z) \propto f_{(R,Z)}( r ,z)\cdot\mathbf{1}_{( r ,z)\in E^*}
=f_R( r ) f_Z(z)\cdot\mathbf{1}_{( r ,z)\in E^*},\]
and therefore the conditional density of $R$ is given by
\[f_{R\mid Z; (R,Z)\in E^*}( r \mid z) \propto f_R( r )\cdot\mathbf{1}_{( r ,Z)\in E^*}.\]
Returning to our definitions, we see that,  letting $\textnormal{Cluster}^* =  \textnormal{Cluster}(X)$,
\[ ( r ,Z)\in E^*  \ \Leftrightarrow \  r  \in  \mathcal{S}' ,\]
and therefore, we have calculated that the distribution of $R$
conditional on $Z$ and on $\textnormal{Cluster}(X)$
is 
\[f_{R\mid Z; (R,Z)\in E^*}( r \mid z) \propto f_R( r )\cdot\mathbf{1}_{ r  \in\mathcal{S}'}.\]
This proves the desired result about the distribution of $R$.
The validity of the p-value $P'$ follows as an immediate consequence.

\subsection{Proof of Proposition~\ref{prop:trunc_relate}}
Let $v = \frac{\mathbf{1}_{C_1}}{|C_1|} - \frac{\mathbf{1}_{C_2}}{|C_2|}$, 
specializing the construction from before to the case $K=2$ (i.e., we are comparing clusters $k=1$ and $k'=2$). Define
\[\phi = \sqrt{\frac{ r }{m-2 }} \cdot \|v\|_2 \cdot \|\mathcal{P}_1 X\|_F.\]
We can rearrange terms to obtain
\[ \frac{r}{m-2}  = \frac{\phi^2}{\|v\|^2_2 \cdot\|\mathcal{P}_1 X\|^2_F}\]
and therefore
\[\frac{r}{m-2+r} = \frac{ \frac{r}{m-2}}{1 + \frac{r}{m-2}}  = \frac{\frac{\phi^2}{\|v\|^2_2 \cdot\|\mathcal{P}_1 X\|^2_F}}{1+\frac{\phi^2}{\|v\|^2_2 \cdot\|\mathcal{P}_1 X\|^2_F}} = \frac{\phi^2}{\phi^2 + \|v\|^2_2 \cdot\|\mathcal{P}_1 X\|^2_F}.\]
We  compute
\begin{align*}
x'( r ) 
&= \left(\sqrt{\frac{r}{m-2+r}} \cdot \frac{\mathcal{P}_0 X}{\|\mathcal{P}_0 X\|_F}+\sqrt{\frac{m-2}{m-2+ r}}\cdot \frac{\mathcal{P}_1 X}{\|\mathcal{P}_1 X\|_F}  \right)  \cdot \sqrt{ \|\mathcal{P}_0 X\|^2_F +  \|\mathcal{P}_1 X\|^2_F }+ \mathcal{P}_2X\\
&= \left(\phi \cdot \frac{\mathcal{P}_0 X}{\|\mathcal{P}_0 X\|_F}+ \|v\|_2 \cdot  \|\mathcal{P}_1 X\|_F\cdot \frac{\mathcal{P}_1 X}{\|\mathcal{P}_1 X\|_F}  \right)  \cdot \sqrt{\frac{ \|\mathcal{P}_0 X\|^2_F + \|\mathcal{P}_1 X\|^2_F}{\phi^2 + \|v\|^2_2 \cdot\|\mathcal{P}_1 X\|^2_F}}+ \mathcal{P}_2X\\
&=x(\phi) \cdot \|v\|_2 \cdot \sqrt{\frac{ \|\mathcal{P}_0 X\|^2_F + \|\mathcal{P}_1 X\|^2_F}{\phi^2 + \|v\|^2_2 \cdot\|\mathcal{P}_1 X\|^2_F}}- \mathcal{P}_2X \cdot\left( \|v\|_2\cdot \sqrt{\frac{ \|\mathcal{P}_0 X\|^2_F + \|\mathcal{P}_1 X\|^2_F}{\phi^2 + \|v\|^2_2 \cdot\|\mathcal{P}_1 X\|^2_F}}  - 1\right),
\end{align*}
where for the last step we apply the calculation~\eqref{eqn:x(phi)_in_our_notation}.
Furthermore, in the case $K=2$, we can verify that $\mathcal{P}_2 = \frac{\mathbf{1}_n\mathbf{1}_n^\top}{n}$, the projection to the span of $\mathbf{1}_n$.
Therefore, we can write
\[x'( r )  = a\cdot x(\phi) + \mathbf{1}_n\cdot  b^\top,\]
where 
\[a = \|v\|_2\cdot\sqrt{\frac{ \|\mathcal{P}_0 X\|^2_F + \|\mathcal{P}_1 X\|^2_F}{\phi^2 + \|v\|^2_2 \cdot\|\mathcal{P}_1 X\|^2_F}} \]
and 
\[b =- \left( \|v\|_2 \cdot \sqrt{\frac{ \|\mathcal{P}_0 X\|^2_F + \|\mathcal{P}_1 X\|^2_F}{\phi^2 + \|v\|^2_2 \cdot\|\mathcal{P}_1 X\|^2_F}} - 1\right) \cdot X^\top \frac{\mathbf{1}_n}{n}.\]
By our assumption on the clustering procedure, we therefore have
\[\textnormal{Cluster}(x'( r ))  =\textnormal{Cluster}(x(\phi)) ,\]
and so $ r \in\mathcal{S}'$ if and only if $\phi\in\mathcal{S}$, as desired.

\subsection{Implementation details}
In our experiments, the three variants of \citet{gao2020selective}'s method (i.e., using the true $\sigma$,
or the estimates $\hat\sigma_{\textnormal{all}}$ or $\hat\sigma_{\textnormal{clustered}}$) are run using 
the code provided by \citet{gao2020selective} in the R package \texttt{clusterpval}. We now give details for implementation of our proposed method.

\subsubsection{Notes on computation in the $K=2$ case}
Computing the p-value 
\[P'=1-F_{ \textnormal{F}_{q,(m-2)q}}(R;\mathcal{S}')\]
for our method often requires extremely precise calculations of the CDF of the F distribution due to the truncation.

 To estimate the CDF in the tail of the F distribution, we use  \citet{li2002approximation}'s approximation,
\[ F_{\textnormal{F}_{k,\ell}}(t) \approx  F_{\chi^2_k}\left(\frac{2\ell + \frac{kt}{3} +k-2}{2\ell + \frac{4kt}{3}}\cdot  kt\right).\]
Note that \citet{li2002approximation}'s method is accurate in the regime where $k$ is finite while $\ell\rightarrow\infty$, which 
is appropriate to our setting as we apply the approximation with $k=q$ and $\ell=(m-2)q$.
Combining all these calculations means that we can approximate $P'$ via the CDF of a truncated $\chi^2$ distribution,
\[P'\approx 1-F_{\chi^2_q}\left(\frac{2(m-2)q + \frac{qR}{3} +q-2}{2(m-2)q + \frac{4qR}{3}}\cdot qR \, ; {\tilde{\mathcal{S}}}'\right)\]
where 
\[ \tilde{\mathcal{S}}' = \left\{\frac{2(m-2)q + \frac{qr}{3} +q-2}{2(m-2)q +\frac{4qr}{3}}\cdot qr: r\in \mathcal{S}'\right\}.\]
Finally, we use the \texttt{TChisqRatioApprox} function from \citet{gao2020selective}'s R package \texttt{clusterpval} for this remaining calculation with the truncated $\chi^2$ distribution.

\subsubsection{Importance sampling algorithm for the $K\geq2$ case}\label{app:importance_sampling}

For the setting where $K>2$, or where $K=2$ but we do not have an exact characterization of the truncation set $\mathcal{S}'$,
we use importance sampling to approximate the p-value  $P'$. 

In order to run importance sampling, it is convenient for us to transform the F distribution to a Beta distribution, since the Beta
distribution takes values over a bounded range.
From properties of these two distributions, it holds that $Z\sim \textnormal{Beta}(a/2,b/2)$ if and only if $\frac{b}{a}\cdot\frac{Z}{1-Z}\sim \textnormal{F}_{a,b}$,
or equivalently, $R\sim F_{a,b}$ if and only if $\frac{R}{\frac{b}{a}+R}\sim \textnormal{Beta}(a/2,b/2)$.
We can then derive an analogous transformation for the \emph{truncated} versions of these distributions:
for any $r$ it holds that
\[F_{\textnormal{F}_{q,(m-2)q}}\left(r;\mathcal{S}'\right) = 
F_{\textnormal{Beta}(q/2,(m-2)q/2)}\left(\frac{r}{m-2+r};\mathcal{S}''\right),\]
where 
$\mathcal{S}'' = \{z\in(0,1) : (m-2)\frac{z}{1-z} \in\mathcal{S}'\}$.
The p-value $P'$ is therefore equal to
\[P' = 1 - F_{\textnormal{F}_{q,(m-2)q}}(R;\mathcal{S}') = 1 - F_{\textnormal{Beta}(q/2,(m-2)q/2)}\left(\frac{R}{m-2+R};\mathcal{S}''\right).\]

In order to calculate this p-value, we need to be able to perform a calculation of the form
 \begin{multline*}1-F_{\textnormal{Beta}(q/2,(m-2)q/2)}\left(z;\mathcal{S}''\right)
  = \Ppst{Z\sim \textnormal{Beta}(q/2,(m-2)q/2)}{{Z>z}}{Z\in\mathcal{S}''}\\
  = \frac{\Pp{Z\sim \textnormal{Beta}(q/2,(m-2)q/2)}{Z>z, \ Z\in\mathcal{S}''}}{\Pp{Z\sim \textnormal{Beta}(q/2,(m-2)q/2)}{Z\in\mathcal{S}''}}, \end{multline*}
and then apply this calculation at the value $z=\frac{R}{m-2+R}$.
We estimate the numerator and denominator simultaneously, using importance sampling
with the proposal distribution
\[\textnormal{TN}\left(\frac{R}{m-2+R},\alpha^2;0,1\right),\]
which is the truncated normal distribution---i.e., the $\textnormal{N}\left(\frac{R}{m-2+R},\alpha^2\right)$ distribution truncated to the interval $[0,1]$.

Our procedure is:
\begin{itemize}
\item Draw $Z^{(1)},\dots,Z^{(N)}\iidsim  \textnormal{TN}\left(\frac{R}{m-2+R},\alpha^2;0,1\right)$, for $N$ draws.
\item Compute importance weights
\[w(Z^{(i)}) = \frac{f_{\textnormal{Beta}(q/2,(m-2)q/2)}(Z^{(i)})}{f_{\textnormal{TN}\left(\frac{R}{m-2+R},\alpha^2;0,1\right)}(Z^{(i)})}\textnormal{ for }i\in[N],\]
where $f_{\textnormal{Beta}(q/2,(m-2)q/2)}$ and $f_{\textnormal{TN}\left(\frac{R}{m-2+R},\alpha^2;0,1\right)}$ denote the densities of the
respective distributions.
\item Estimate
\[P' \approx \frac{\sum_{i=1}^N w(Z^{(i)})\cdot \mathbf{1}\left\{ Z^{(i)} > \frac{R}{m-2+R}, Z^{(i)}\in \mathcal{S}'' \right\}}{\sum_{i=1}^Nw(Z^{(i)})\cdot  \mathbf{1}\left\{ Z^{(i)}\in \mathcal{S}'' \right\}}.\]
\end{itemize}
The  tuning parameter  $\alpha$ is adjusted based on empirical performance---specifically, we choose $\alpha$ so that $\frac{1}{N} \sum_{i=1}^N \mathbf{1}\{Z^{(i)}\in \mathcal{S}''\}\approx 0.5$,
to ensure that our proposal distribution $ \textnormal{TN}\left(\frac{R}{m-2+R},\alpha^2;0,1\right)$ is neither too concentrated nor too dispersed 
to accurately approximate the truncated target distribution.

\end{document}